\documentclass[a4paper,10pt]{article}
\usepackage{amsmath}
\usepackage{amssymb}
\usepackage{color}

\newcommand{\al}{\alpha}
\newcommand{\pa}{\partial}

\newcommand\be{\begin{equation}}
\newcommand\ee{\end{equation}}
\newcommand\nn{\nonumber}

\begin{document}

\vskip 12mm

\begin{center} 
{\Large \bf  Interactions of Irregular Gaiotto States in Liouville Theory}
\vskip 10mm
{ \large  Sang-Kwan Choi$^{a}$, Dimitri Polyakov$^{a,b,}$\footnote{email:polyakov@scu.edu.cn;polyakov@sogang.ac.kr},
Cong Zhang$^{a}$

\vskip 8mm
$^{a}$ {\it  Center for Theoretical Physics, College of Physical Science and Technology}\\
{\it  Sichuan University, Chengdu 6100064, China}\\

\vskip 2mm

$^{b}$ {\it Institute of Information Transmission Problems (IITP)}\\
{\it  Bolshoi Karetny per. 19/1, Moscow 127994, Russia}\\

}
\end{center}

\vskip 15mm

\begin{abstract}

We compute the correlation functions of irregular Gaiotto states appearing in
the colliding limit of the Liouville theory 
by using ``regularizing'' conformal transformations mapping the irregular
(coherent) states to regular vertex operators in the Liouville theory.
The $N$-point correlation functions of the irregular vertex operators of arbitrary ranks are expressed
in terms of $N$-point correlators of primary fields times the factor that involves
regularized higher-rank Schwarzians of the above conformal transformation.
In particular, in the case of three-point functions the general answer is expressed in terms of 
DOZZ (Dorn-Otto-Zamolodchikov-Zamolodchikov) structure constants times 
exponents of regularized higher-derivative Schwarzians.
The explicit examples of the regularization  are given 
for the ranks one and two.

\end{abstract}

\vskip 12mm

\setcounter{footnote}{0}

\section{\bf Introduction} 
Irregular (coherent) Gaiotto states emerge in Liouville field theory
in the colliding limit, relevant to  extensions of the AGT conjecture
to Argyres-Douglas type of gauge theories with asymptotic freedom 
\cite{AGT, AD, GT_2012, G_2009, Whittaker, CRZ_2014, RP}.
The irregular states of rank $N$ are the simultaneous eigenstates of $N+1$
Virasoro generators:
\begin{eqnarray}
L_n |U_N \rangle = \rho_N(n)|U_N \rangle
\nonumber \\
N\leq{n}\leq{2N}
\end{eqnarray}
and are annihilated by higher positive Virasoro generators
($n>2N$). This generalizes the definition of primary fields
(which technically have rank zero).
Just as the regular vertex operators for primary (rank zero)
fields in Liouville theory
can  be expressed as $V_\alpha=:e^{\alpha_0\phi}:$ (where $\alpha_0$ can be 
regarded as ``electric'' charge), the irregular vertex operators for
rank $N$ coherent states can be constructed as
\begin{eqnarray}
|U_N \rangle=U_N|0 \rangle
\nonumber \\
U_N=:e^{\sum_{n=0}^N\alpha_n\partial^n\phi}:
\end{eqnarray}
where $\alpha_1,\alpha_2...$ correspond to coefficients
of the multipole expansion and are related to the geometry of the region
where the colliding operators are located (e.g.
 with $\alpha_1$ being a characteristic size of the region).
These coefficients are related to Virasoro eigenvalues $\rho_n(N)$
according to
\begin{eqnarray}
\rho_n(N)={1\over2}\sum_{n=0}^{N}\alpha_n\alpha_{N-n}
\end{eqnarray}
Computing correlation functions describing interactions of
the irregular states in Liouville theory is known to be a hard and
tedious problem, especially beyond two-point functions and rank one case
\cite{EG_2009, GT_2012, MMM_0909, CRZ_2014, CRZ_2015}.
 In this work we address this problem by using 
the conformal transformations
that maps operators for coherent 
states into regular vertex operators, 
expressing the interactions of the irregular states in terms of regular 
correlators in Liouville theory. In particular, since the structure constants 
of the Liouville theory are known, this 
conformal transformation  makes it possible
 to express the cubic interactions of the irregular states of arbitrary
rank in terms 
of Dorn-Zamolodchikov correlators in Liouville theory. The final formula
for the correlators, however, is complicated, since it involves the
generalized higher-derivative Schwarzians of the conformal transformation that
are singular at the insertion points of the correlators and have to be 
regularized in a rather tedious way. In our work, we limit ourselves 
performing this regularization for the case of 3-point functions of the rank
two irregular  vertex operators. The rest of this paper is organized as 
follows.
In the Section 2, we study the behavior of correlators under the conformal
 transformation mapping the irregular blocks into regular and derive the
general result expressing $N$-point correlators of irregular vertices of arbitrary ranks in terms of regular conformal blocks. The answer involves 
the higher-derivative generalized Schwarzians of this conformal transformation
that need to be regularized. In the Appendix section
we  explicitly perform such a regularization for three-point correlators
of the rank two states. In the concluding section we discuss the implications 
of our results.

\section{\bf Conformal Map: Irregular to Regular States}

Consider a rank p irregular vertex operator:
\begin{eqnarray}
W_p(\alpha_0,\alpha_1,...,\alpha_p|\xi)=:e^{\sum_{n=0}^p\alpha_n\partial^n\phi}:(\xi)
\end{eqnarray}
where $\phi$ is Liouville field and the $N$-point correlators
\begin{eqnarray}
A_N=\Big\langle \prod_{j=1}^{N}W_{p_j}(\alpha_0^{(j)},...,\alpha_{p_j}^{(j)}|\xi_j) \Big\rangle
\end{eqnarray}
Consider conformal transformation:
\begin{eqnarray}
f(z)=e^{-i\sum_{j=1}^N(z-\xi_j-i\epsilon)^{-1}}
\end{eqnarray}
with the small $\epsilon$ parameter introduced in order to control regularizations.
This transformations maps the half-plane to compact Riemann surface 
with all the the points $\xi_1,...\xi_N$ (originally located on the real line)
are glued together at zero for $\epsilon=0$ (and are infinitely close to each other
if $\epsilon$ is nonzero). Gluing points together is the trick that will be used below to compare correlation
functions before and after conformal transformations. 
First of all, let us check how the conformal 
transformation (6) acts on individual irregular vertices.
It is straightforward to check that the 
transformation  (6) maps irregular vertex operators (4) into regular.
To see this, first consider the infinitezimal transformations of $W_p$.
We have
\begin{eqnarray}
\delta_\epsilon{W_p}(\xi)={1\over2}\lbrack\oint{{dz}\over{2i\pi}}\epsilon(z):\partial\phi\partial\phi:(z); W_p(\xi)\rbrack
\nonumber \\
=\sum_{n=0}^p\alpha_n\partial^n(\epsilon\partial\phi)W_p(\xi)+{1\over2}\sum_{n_1,n_2=0}^p
{{n_1!n_2!}\over{(n_1+n_2+1)!}}\alpha_{n_1}\alpha_{n_2}\partial^{n_1+n_2+1}\epsilon{W_p}(\xi)
\end{eqnarray}
 This infinitezimal relation is straightforward to integrate 
(e.g. by imposing a composition constraint). The integrated form of (7) for
finite conformal transformations $z\rightarrow{f(z)}$ is then given by
\begin{eqnarray}
W_p(z)\rightarrow{\tilde{W_p}}(f(z))=
\mathrm{exp} \bigg\{\alpha_0\phi+\sum_{n=1}^p\sum_{q=1}^nB_{n|q}(f(z);z)\alpha_n\partial^q\phi(f(z))
\nonumber\\
+\sum_{n_1,n_2=0}^{p}n_1! n_2! \, \alpha_{n_1} \alpha_{n_2} S_{n_1|n_2}(f(z);z) \bigg\}
\end{eqnarray}
Here 
\begin{eqnarray}
B_{n|q}(f(z);z)=\sum_{n|n_1...n_q}{{\partial^{n_1}\phi...\partial^{n_q}\phi}\over{n_1!...n_q!r(n_1)!...r(n_q)!}}
\end{eqnarray}
are the restricted length $q$ Bell polynomials in derivatives of $f$, with the sum taken over
the ordered length $q$  partitions of $n=n_1+...+n_q$; with $0<n_1\leq{n_2}....\leq{n_q}$
anq $r(n_i)$ is  multiplicity of element $n_i$ in the partition
(e.g. for $8=2+3+3$ $q(1)=0,q(2)=1$ and $q(3)=2$).
Next, $S_{n_1|n_2}$ are the generalized rank $(n_1,n_2)$ Schwarzians of the conformal transformation $f(z)$, defined
according to \cite{SF}:
\begin{eqnarray}
S_{n_1|n_2}(f;z)=
{1\over{n_1!n_2!}}\sum_{k_1=1}^{n_1}\sum_{k_2=1}^{n_2}
\sum_{m_1\geq{0}}\sum_{m_2\geq{0}}\sum_{p\geq{0}}
\sum_{q=1}^p
(-1)^{k_1+m_2+q}2^{-m_1-m_2}(k_1+k_2-1)!
\nonumber \\
\times
{{\partial^{m_1}B_{n_1|k_1}(f(z);z)\partial^{m_2}B_{n_2|k_2}(f(z);z)
B_{p|q}(g_1,...,g_{p-q+1})}
\over{m_1!m_2!p!(f^\prime(z))^{k_1+k_2}}}
\nonumber \\
g_s=2^{-s-1}(1+(-1)^s){{{{d^{s+1}f}\over{dz^{s+1}}}}\over{(s+1)f^\prime(z)}};
s=1,...,p-q+1
\end{eqnarray}
for $n_1,n_2\neq{0}$
with the sum over the non-negative numbers $m_1,m_2$ and $p$ 
taken over all the combinations satisfying
$$m_1+m_2+p=k_1+k_2$$
Also, $S_{0|0}(f;z)=ln(f^\prime(z))$ ,
$S_{1|0}=S_{0|1}={{f^{\prime\prime}(z)}\over{2f^\prime(z)}}$
and $S_{1|1}$ is a usual Schwarzian derivative (up to conventional factor of ${1\over6}$).
As $z\rightarrow{\xi_j}$, the coefficients in front of 
derivatives $\partial^q\phi$ ($q\neq{0}$)
in the exponent of ${\tilde{W}}_p(f(z))$, determined by the length $q$ Bell polynomials
in $f$, are damped exponentially as $\sim{e^{-{q\over{\epsilon}}}}$, so only the regular part
$\sim{\alpha_0\phi}$ survives and the operators become regular in new coordinates.
On the other hand, the price paid for the regularity is the appearance of
the generalized Schwarzians $S_{n_1|n_2}(f;z)$ in the transformation law 
for the irregular vertices. All  of these Schwarzians  have inverse power behavior in $\epsilon$
and must be regularized as $\epsilon\rightarrow{0}$.
The final step is to compute the overlap  deformation resulting from contractions
of $T(z)={1\over{2}}:\partial\phi\partial\phi:$ with different vertex operators
(i.e. each of $\partial\phi$'s contracting with different vertex) and to integrate it.
This altogether is equivalent to integrating the Ward identities and, in the limit $\epsilon\rightarrow{0}$,
the integrated overlap deformation determines the difference between the correlators computed
on the half-plane and on the Riemann surface defined by the conformal map $f(z)$, with the transformation
laws (8) for the vertex operators.
The relevant infinitezimal  overlap transformation of the $N$-point correlator is given by
\begin{eqnarray}
\delta_{overlap}A_N(\xi_1,...\xi_N)
=A_N(\xi_1,...\xi_N)\sum_{j=1}^{N-1}\sum_{k=j+1}^N\sum_{n_j=1}^{p_j}\sum_{n_k=1}^{p_k}
\alpha_{n_j}^{(j)}\alpha_{n_k}^{(k)}
\nonumber \\
\times \left(n_k!\partial^{n_j}_{\xi_j}{{\epsilon(\xi_j)}\over{(\xi_j-\xi_k)^{n_k+1}}}
+n_j!\partial^{n_k}_{\xi_k}{{\epsilon(\xi_k)}\over{(\xi_k-\xi_j)^{n_j+1}}}\right)
\end{eqnarray}

It is straightforward to integrate it to obtain the contribution of the overlap to the  deformation of 
$A_N$ under the finite conformal transformation $z\rightarrow{f(z)}$.
The result is

\begin{eqnarray}
A_N(\xi_1,...,\xi_N)\rightarrow
A_N(f(\xi_1),...,f(\xi_N))exp \bigg\{ \sum_{j=1}^{N-1}\sum_{k=j+1}^N\sum_{n_j=1}^{p_j}\sum_{n_k=1}^{p_k}
\alpha_{n_j}^{(j)}\alpha_{n_k}^{(k)}
\nonumber \\
\times
\Big(\sum_{{p_j}=0}^{n_j-1}\sum_{{p_k}=0}^{n_k}
\Big(
{{B_{n_j|n_j-p_j}(f(\xi_j);\xi_j)B_{n_k|n_k-p_k}(f(\xi_j);\xi_j)}\over{(f(\xi_j)-f(\xi_k))^{n_j+n_k-p_j-p_k}}}-
{1\over{(\xi_j-\xi_k)^{n_j+n_k-p_j-p_k}}}\Big)\Big)\bigg \}
\end{eqnarray}

The relation between the correlators
of the irregular and regular states is given by the transformation
law (8) divided by the overlap deformation (12).
Thus for $N$-point correlator of irregular states of arbitrary rank one has:

\begin{eqnarray}
A_N(\xi_1,...,\xi_N)=S_N(\xi_1,...,\xi_N)
\nonumber \\
\times
exp \bigg \{{1\over2}\sum_{j=1}^N\sum_{{k=1}^{p_j}}\sum_{l=1}^{p_j}
\alpha_k^{(j)}\alpha_l^{(j)}S_{k|l}(f(z);z)|_{z=\xi_j}
\nonumber \\
-\sum_{j=1}^{N-1}\sum_{k=j+1}^N\sum_{n_j=1}^{p_j}
\sum_{n_k=1}^{p_k}                           
\alpha_{n_j}^{(j)}\alpha_{n_k}^{(k)}
\nonumber \\ 
\times
\Big(\sum_{{p_j}=0}^{n_j-1}\sum_{{p_k}=0}^{n_k}\Big({{B_{n_j|n_j-p_j}(f(\xi_j);\xi_j)
B_{n_k|n_k-p_k}(f(\xi_j);\xi_j)}\over{(f(\xi_j)-f(\xi_k))^{n_j+n_k-p_j-p_k}}}
-{1\over{(\xi_j-\xi_k)^{n_j+n_k-p_j-p_k}}}\Big)\Big) \bigg\}
\end{eqnarray}
where
$S_N$ is the $N$-point correlator of regular Liouville primaries:

\begin{equation}
S_N(\xi_1,...,\xi_N)= \big \langle e^{\alpha_0^{(1)}\phi}(\xi_1)...e^{\alpha_0^{(N)}\phi}(\xi_N) \big \rangle
\end{equation}

This reduces the problem of describing the interactions of irregular 
states of arbitrary rank
in terms of those of the regular states.
In particular, for $N=3$, $S_3$ is well-known and related to the regular 
Liouville
S-matrix \cite{DO, ZZ}. However, the Schwarzians and overlap factors 
(involving  sums over the
restricted Bell polynomials of the conformal transformation (6)) 
are singular at the insertion points
of the irregular vertex operators and need to be regularized. In the 
Appendix section,
we show how to perform such a regularization explicitly for
the rank two irregular states ($p_1=p_2=p_3=2$) and for $N=3$. 
It already turns to be quite cumbersome.
 The generalizations of this regularization procedure
 for higher number of points 
and for higher ranks are in principle straightforward but require 
 far more tedious calculations.

\section{Conclusions}

In this work, we analyzed interactions of the irregular states in Liouville
theory by using conformal transformations that map the irregular vertex
operators into regular. In particular, this allows to express three-point 
functions of irregulars of arbitrary rank in terms of Liouville structure 
constants given
by DOZZ (Dorn-Otto-Zamolodchikov-Zamolodchikov) formula.
The price one pays
is the appearance of the objects, such as
 higher-derivative Schwarzians and overlap factors (sums over restricted Bell
 polynomials), in the  transformation law for the vertices.
They are singular at the insertion points and  need to be 
regularized by a rather cumbersome procedure.
Interestingly, all these objects appear in the solution
of the well-known number theory problem of finding the closed analytic
expressions for numbers of restricted partitions \cite{SF}.
In the current paper, we restrict ourselves to the maximum rank two
and the three-point function.
In our future work (currently in progress), we hope to be able to develop the
algorithm which simplifies the regularization scheme and can be applied
to analyze the higher-rank interactions.
Irregular states, apart from being relevant to AGT conjecture
(extended to Argyres-Douglas class of gauge theories) 
also appear in the interplay between open string field theory 
and higher-spin gauge theories, as the irregular vertex operators
can be understood as generating wavefunctions for higher-spin
vertex operators in bosonic string theory.
Given certain constraints  on $\alpha_n^{(m)}$, the irregular vertices
of the lower ranks (1 and 2)
form non-trivial solutions of open string field theory (OSFT) equations of 
motions, describing certain  special  limits of
collective higher-spin configurations. Using the formalism developed
in this paper we hope to extend these particular solutions to arbitrary ranks
in order to describe the OSFT solutions
describing general collective higher-spin vacua in string theory.
These solutions, in general, are parametrized by nontrivial number theory
identities, particularly involving higher-order Schwarzians and Bell 
polynomials. We hope that classifying these solutions may be useful
to elucidate deep underlying relations, existing between number theory, 
bosonic strings and higher spin gauge theories. 
    
\section{Appendix}
In this  section, we perform explicit regularizations 
of the higher-derivative Schwarzians and the overlap factors,
for the case of rank 2 irregular states. We limit ourselves
to 3-point functions.
\section*{4a. Transformation}
For each irregular vertex operator of rank 2,
the conformal transformation  law is
\be
\begin{split}
e^{\al_0 \phi+\al_1 \pa \phi+\al_2 \pa^2 \phi}
\stackrel{z \to f(z)}{\longrightarrow} ~
&\mathrm{exp} \Big[ \al_0 \phi+\al_1 B_{1|1} \pa \phi +\al_2 
\left(B_{2|1} \pa \phi+B_{2|2} \pa^2 \phi \right) +\frac{1}2 \al_0^2 S_{0|0}
\\
&+\al_0 \al_1 S_{1|0}
+2 \al_0 \al_2 S_{2|0}+\frac12 \al_1^2 S_{1|1}
+2 \al_1 \al_2 S_{2|1}+2\al_2^2 S_{2|2}\Big]
\end{split}
\ee
where
\begin{align}
&S_{0|0}=\log f' \,, \qquad  S_{1|0}=\frac12 \frac{f''}{f'} \,,
\qquad S_{1|1}=\frac16 \left( \frac{f''}{f'} \right)'-\frac1{12}
\left(\frac{f''}{f'} \right)^2 \,, 
\nn \\
&S_{2|0}=\frac16\frac{f'''}{f'}-\frac18 \left( \frac{f''}{f'} \right)^2 \,,
\qquad
S_{2|1}=\frac18\left(\frac{f''}{f'} \right)^3-\frac16 \frac{f'' \, f'''}{(f')^2}
+\frac1{24} \frac{f^{(4)}}{f'} \,,
\nn \\
&S_{2|2}=\frac1{120}\frac{f^{(5)}}{f'}
-\frac1{24}\frac{f^{(4)} \, f''}{(f')^2}-\frac1{24}\left(\frac{f'''}{f'} \right)^2
+\frac16 \frac{(f'')^2 f'''}{(f')^3}-\frac3{32} \left(\frac{f''}{f'} \right)^4 \nn
\end{align}
Next, consider the overlap transformation to the 3-point correlator.
The overlap contribution between $(V(\xi_1),V(\xi_2))$ is given below,
the overlaps between $V(\xi_1),V(\xi_2)$ and $V(\xi_2),V(\xi_3)$
are obtained similarly, by replacing variables $\alpha,\beta,\xi$
accordingly.
Denote $f_1=f(\xi_1)$ and $f_2=f(\xi_2)$. We have:
\be
\begin{split}
\mathrm{overlap}&
\left(e^{\al_0 \phi+\al_1 \pa \phi+\al_2 \pa^2 \phi}(\xi_1)
e^{\beta_0 \phi+\beta_1 \pa \phi+\beta_2 \pa^2 \phi} (\xi_2) \right)
\\
\stackrel{z \to f(z)}{\longrightarrow}
\mathrm{exp} &\Bigg[ -\al_0 \beta_0 \log (f_1-f_2)
-\al_1 \beta_0 \frac{f_1'}{f_1-f_2}
+\al_0 \beta_1 \frac{f_2'}{f_1-f_2}
-\al_1 \beta_1 \frac{f_1' \, f_2'}{(f_1-f_2)^2}
\\
&-\al_2 \beta_0 \left( \frac{f_1''}{f_1-f_2}
-\left(\frac{f_1'}{f_1-f_2}\right)^2 \right)
+\al_0 \beta_2 \left( \frac{f_2''}{f_1-f_2}
+\left(\frac{f_2'}{f_1-f_2}\right)^2 \right)
\\
&-\al_2 \beta_1 \left( \frac{f_1'' \, f_2'}{(f_1-f_2)^2}
-\frac{2(f_1')^2 f_2'}{(f_1-f_2)^3} \right)
-\al_1 \beta_2 \left( \frac{f_1' \, f_2''}{(f_1-f_2)^2}
+\frac{2f_1' (f_2')^2}{(f_1-f_2)^3} \right)
\\
&-\al_2 \beta_2 \left( \frac{f_1'' \, f_2''}{(f_1-f_2)^2}
-\frac{2(f_1')^2 f_2''}{(f_1-f_2)^3}
+\frac{2f_1'' (f_2')^2}{(f_1-f_2)^3}
-\frac{6 (f_1')^2 (f_2')^2}{ (f_1-f_2)^4} \right)
\\
&+\al_0 \beta_0 \log (\xi_1-\xi_2)
+\frac{\al_1 \beta_0}{\xi_1-\xi_2}
-\frac{\al_0 \beta_1}{\xi_1-\xi_2}
+\frac{\al_1 \beta_1}{(\xi_1-\xi_2)^2}
-\frac{\al_2 \beta_0}{(\xi_1-\xi_2)^2}
\\
&-\frac{\al_0 \beta_2}{(\xi_1-\xi_2)^2}
-\frac{2 \al_2 \beta_1}{(\xi_1-\xi_2)^3}
+\frac{2\al_1 \beta_2}{(\xi_1-\xi_2)^3}
-\frac{6 \al_2 \beta_2}{(\xi_1-\xi_2)^4}
 \Bigg]
\end{split}
\ee

\section*{4.b Regularization: 2-point function}
We start by regularilizing the 2-point function first. We have:
\be
A_2=\left\langle e^{\al_0 \phi+\al_1 \pa \phi+\al_2 \pa^2 \phi}(\xi_1)
e^{\beta_0 \phi+\beta_1 \pa \phi+\beta_2 \pa^2 \phi} (\xi_2) \right\rangle \,,
~~~~~~~
f(z)=e^{-i \left(\frac{1}{z-\xi_1}+\frac{1}{z-\xi_2}\right)}
\ee

\subsection*{Regularizing generalized Schwarzians:}

Denote $\xi_{12} \equiv \xi_1-\xi_2$.

For $e^{\al_0 \phi+\al_1 \pa \phi+\al_2 \pa^2 \phi}(\xi_1)$,
\begin{align}
&S_{0|0} \sim \log \left(i e^{-i/\xi_{12}} \right) \,,
\qquad S_{1|0} \sim \frac{i}{2 \xi_{12}^2} \,,
\qquad S_{1|1} \sim \frac{7}{12\xi_{12}^4}+\frac1{\xi_{12}^2}  \,,
 \\
&  S_{2|0} \sim -\frac{7}{24 \xi_{12}^4}-\frac{i}{2\xi_{12}^3} \,,
\qquad
S_{2|1} \sim -\frac{1}{4 \xi_{12}^5}-\frac1{\xi_{12}^3} \,,
\qquad
S_{2|2} \sim -\frac{323}{480 \xi_{12}^8}-\frac{19}{12 \xi_{12}^6}
+\frac1{2\xi_{12}^4} \nn
\end{align}

For $e^{\beta_0 \phi+\beta_1 \pa \phi+\beta_2 \pa^2 \phi}(\xi_2)$,
$\xi_{12} \to -\xi_{12}$.

\subsection*{Regularizing the overlap part:}
Denote $X_k\equiv1/\xi_{12}^k$ and $F_1\equiv e^{-i X_1}$.
Here we omit the free correlator part.
Below, one by one,  we present the  regularized coefficients
in front of $\alpha_i\beta_j;i,j=0,1,2$ in the exponential.
$\alpha_0 \beta_0$
\be
-\log \left(F_1-\frac{1}{F_1}\right)
\ee

$\alpha_1 \beta_0$
\be
-\frac{i F_1^2 \left(\left(F_1^2-1\right) X_2+2 i X_3\right)}{\left(F_1^2-1\right)^2}
\ee

$\alpha_0 \beta_1$
\be
\frac{i \left(\left(F_1^2-1\right) X_2+2 i F_1^2 X_3\right)}{\left(F_1^2-1\right)^2}
\ee

$\alpha_1 \beta_1$
\be
\begin{split}
\frac{F_1^2}{\left(F_1^2-1\right)^4}
 &\bigg(4 i \left(F_1^4-1\right) X_2 X_3+\left(F_1^2-1\right)^2 X_2^2 
 \\
 &-2 \left(-i \left(F_1^4-1\right) X_5-3 \left(F_1^2-1\right)^2 X_4+\left(F_1^4+4 F_1^2+1\right) X_3^2\right)\bigg)
\end{split}
\ee

$\alpha_2 \beta_0$
\be
\begin{split}
-\frac{F_1^2}{\left(F_1^2-1\right)^4}
 &\bigg(4 i \left(F_1^4-1\right) X_2 X_3+\left(F_1^2-1\right)^2 X_2^2-2 \Big(-i \left(F_1^4-1\right) X_5
 \\
 &+i \left(F_1^2-1\right)^3 X_3-3 \left(F_1^2-1\right)^2 X_4+\left(F_1^4+4 F_1^2+1\right) X_3^2\Big)\bigg)
\end{split}
\ee

$\alpha_0 \beta_2$
\be
\begin{split}
-\frac1{\left(F_1^2-1\right)^4}
&\bigg(4 i  \left(F_1^4-1\right)F_1^2 X_2 X_3 + \left(F_1^2-1\right)^2 F_1^2 X_2^2-2  \Big(-i \left(F_1^4-1\right) F_1^2 X_5
\\
&+ i \left(F_1^2-1\right)^3 X_3-3 \left(F_1^2-1\right)^2 F_1^2 X_4+ \left(F_1^4+4 F_1^2+F_1\right)F_1^2 X_3^2  \Big) \bigg)
\end{split}
\ee

$\alpha_2 \beta_1$
\begin{align}
&-\frac{F_1^2}{3 \left(F_1^2-1\right)^6}
\Bigg[3 i \left(F_1^2+1\right) \left(F_1^2-1\right)^3 X_2^3+24 i \left(F_1^2+1\right) \left(F_1^2-1\right)^3 X_3^2
\nn \\
&-18 \left(F_1^4+4 F_1^2+1\right) \left(F_1^2-1\right)^2 X_2^2 X_3+3 \left(F_1^2-1\right)^2 \Big(15 i \left(F_1^4-1\right) X_6+20 \left(F_1^2-1\right)^2 X_5
\nn \\
&-2 \left(F_1^4+4 F_1^2+1\right) X_7\Big)+6 \left(F_1^2-1\right) X_2 \Big(\left(F_1^2-1\right)^3 X_3+3 i \left(F_1^2-1\right) \big(3 \left(F_1^4-1\right) X_4
\nn \\
&+i \left(F_1^4+4 F_1^2+1\right) X_5\big)-3 i \left(F_1^6+11 F_1^4+11 F_1^2+1\right) X_3^2\Big)
\\
&-6 \left(F_1^2-1\right) X_3 \left(9 \left(F_1^6+3 F_1^4-3 F_1^2-1\right) X_4+2 i \left(F_1^6+11 F_1^4+11 F_1^2+1\right) X_5\right)
\nn \\
&+4 \left(F_1^8+26 F_1^6+66 F_1^4+26 F_1^2+1\right) X_3^3\Bigg]
\nn
\end{align}

$\alpha_1 \beta_2$
\begin{align}
&-\frac{F_1^2}{3 \left(F_1^2-1\right)^6}
\Bigg[-3 i \left(F_1^2+1\right) \left(F_1^2-1\right)^3 X_2^3-24 i \left(F_1^2+1\right) \left(F_1^2-1\right)^3 X_3^2
\nn\\
&+18 \left(F_1^4+4 F_1^2+1\right) \left(F_1^2-1\right)^2 X_2^2 X_3-3 \left(F_1^2-1\right)^2 \Big(15 i \left(F_1^4-1\right) X_6+20 \left(F_1^2-1\right)^2 X_5
\nn \\
&-2 \left(F_1^4+4 F_1^2+1\right) X_7\Big)+6 \left(F_1^2-1\right) X_2 
\Big(-\left(F_1^2-1\right)^3 X_3+3 \left(F_1^2-1\right) \big(
-3 i \left(F_1^4-1\right) X_4
\nn \\
&+\left(F_1^4+4 F_1^2+1\right) X_5\big)+3 i \left(F_1^6+11 F_1^4+11 F_1^2+1\right) X_3^2\Big)
\nn \\
&+6 \left(F_1^2-1\right) X_3 \left(9 \left(F_1^6+3 F_1^4-3 F_1^2-1\right) X_4+2 i \left(F_1^6+11 F_1^4+11 F_1^2+1\right) X_5\right)
\nn \\
&-4 \left(F_1^8+26 F_1^6+66 F_1^4+26 F_1^2+1\right) X_3^3\Bigg]
\end{align}

$\alpha_2 \beta_2$
\begin{align}
&-\frac{F_1^2}{3 \left(F_1^2-1\right)^8}
\Bigg[-3 \left(F_1^2-1\right)^4 \left(F_1^4+4 F_1^2+1\right) X_2^4-24 i \left(F_1^2-1\right)^3 \left(F_1^6+11 F_1^4+11 F_1^2+1\right) X_2^3 X_3
\nn \\
&+12 \left(F_1^2-1\right)^2 X_2^2 \Big(i \left(F_1^2+1\right) \left(F_1^2-1\right)^3 X_3-3 \left(F_1^2-1\right) \big(3 \left(F_1^6+3 F_1^4-3 F_1^2-1\right) X_4
\nn \\
&+i \left(F_1^6+11 F_1^4+11 F_1^2+1\right) X_5\big)+3 \left(F_1^8+26 F_1^6+66 F_1^4+26 F_1^2+1\right) X_3^2\Big)
\nn \\
&+4 i X_2 \bigg(24 i \left(F_1^4+4 F_1^2+1\right) \left(F_1^2-1\right)^4 X_3^2+3 \left(F_1^2-1\right)^3 \big(20 \left(F_1^2+1\right) \left(F_1^2-1\right)^2 X_5
\nn \\
&+i \left(15 \left(F_1^6+3 F_1^4-3 F_1^2-1\right) X_6+2 i \left(F_1^6+11 F_1^4+11 F_1^2+1\right) X_7\right)\big)
\nn \\
&-6 \left(F_1^2-1\right)^2 X_3 \left(9 \left(F_1^8+10 F_1^6-10 F_1^2-1\right) X_4+2 i \left(F_1^8+26 F_1^6+66 F_1^4+26 F_1^2+1\right) X_5\right)
\nn \\
&+4 \left(F_1^{12}+56 F_1^{10}+245 F_1^8-245 F_1^4-56 F_1^2-1\right) X_3^3\bigg)-2 \Bigg(28 i \left(F_1^2-1\right)^3 \big(F_1^6+11 F_1^4
\nn \\
&+11 F_1^2+1\big) X_3^3+6 \left(F_1^2-1\right)^2 X_3 \Big(10 i F_1^8 X_6-F_1^8 X_7+100 i F_1^6 X_6-26 F_1^6 X_7-66 F_1^4 X_7
\nn \\
&-18 i \left(F_1^2-1\right)^3 \left(F_1^2+1\right) X_4-100 i F_1^2 X_6-26 F_1^2 X_7+26 \left(F_1^2-1\right)^2 \left(F_1^4+4 F_1^2+1\right) X_5
\nn \\
&-10 i X_6-X_7\Big)+3 \left(F_1^2-1\right)^2 \bigg(27 \left(F_1^2-1\right)^2 \left(F_1^4+4 F_1^2+1\right) X_4^2
\nn \\
&+12 i \left(F_1^8+10 F_1^6-10 F_1^2-1\right) X_4 X_5-\left(F_1^2-1\right) \Big(40 \left(F_1^2-1\right)^3 X_6
+i \big(50 \left(F_1^2+1\right) \left(F_1^2-1\right)^2 X_7
\nn \\
&+i \left(14 \left(F_1^6+3 F_1^4-3 F_1^2-1\right) X_8+i \left(F_1^6+11 F_1^4+11 F_1^2+1\right) X_9\right)\big)\Big)
\nn \\
&-\left(F_1^8+26 F_1^6+66 F_1^4+26 F_1^2+1\right) X_5^2\bigg)-6 \left(F_1^2-1\right) X_3^2 \Big(F_1^{10} (6 X_4+i X_5+1)
\nn \\
&+F_1^8 (150 X_4+57 i X_5-5)+2 F_1^6 (120 X_4+151 i X_5+5)-2 F_1^4 (120 X_4-151 i X_5+5)
\nn \\
&+F_1^2 (-150 X_4+57 i X_5+5)-6 X_4+i X_5-1\Big)+\big(F_1^{12}+120 F_1^{10}+1191 F_1^8
\nn \\
&+2416 F_1^6+1191 F_1^4+120 F_1^2+1\big) X_3^4\Bigg)\Bigg]
\end{align}

\section*{4c. Regularization: 3-point function}
Now we generalize the regularization, performed above, 
to the case of the $3$-point forrelator of rank 2 operators.
The correlator and the conformal transformation, in the limit
$\epsilon\rightarrow{0}$, are given by:
\be
\left\langle e^{\al_0 \phi+\al_1 \pa \phi+\al_2 \pa^2 \phi}(\xi_1)
e^{\beta_0 \phi+\beta_1 \pa \phi+\beta_2 \pa^2 \phi} (\xi_2) 
e^{\gamma_0 \phi+\gamma_1 \pa \phi+\gamma_2 \pa^2 \phi} (\xi_3)
\right\rangle \,,
\ee
\be
f(z)=e^{-i \left(\frac{1}{z-\xi_1}+\frac{1}{z-\xi_2}
+\frac{1}{z-\xi_3}\right)}
\ee

\subsection*{Regularization of the generalized Schwarzians:}

Denoting $\xi_{ab} \equiv \xi_a-\xi_b$ and
\be
X_k \equiv \frac1{\xi_{12}^k}+\frac1{\xi_{13}^k}\,, ~~~
Y_k \equiv \frac1{\xi_{21}^k}+\frac1{\xi_{23}^k}\,, ~~~
Z_k \equiv \frac1{\xi_{31}^k}+\frac1{\xi_{32}^k}
\ee
For $e^{\al_0 \phi+\al_1 \pa \phi+\al_2 \pa^2 \phi}(\xi_1)$,
\begin{align}
&S_{0|0} \sim \log \left(i e^{-i X_1} \right) \,,
\qquad 
S_{1|0} \sim \frac{i}2 X_2\,,
\qquad
S_{1|1} \sim \frac{X_4}{2}+\frac{X_2^2}{12}+X_2  \,,
 \nn \\
& S_{2|0} \sim
-\frac{1}{2} \left(\frac{X_4}{2}+\frac{X_2^2}{12}+i X_3\right) \,,
\qquad 
S_{2|1} \sim -\frac{X_5}{6}-\frac{X_2 X_3}{12}-X_3 \,,
\\
&S_{2|2} \sim 
\frac{146 X_8}{480}-\frac{23 X_2 X_6}{60}-\frac{2 X_3 X_5}{5}
-\frac{31 X_4^2}{160}-\frac{20 X_6}{12}+\frac{X_3^2}{12}
+X_4 -\frac{X_2^2}{2}
\nn
\end{align}
For $e^{\beta_0 \phi+\beta_1 \pa \phi+\beta_2 \pa^2 \phi}(\xi_2)$
and $e^{\gamma_0 \phi+\gamma_1 \pa \phi+\gamma_2 \pa^2 \phi} (\xi_3)$,
replace $X_k$ with $Y_k$ and $Z_k$, repectively.

\subsection*{Regularization of the overlap part:}
We show explicitly in the following the overlap contributions between the irregular vertex operators at $\xi_1$ and $\xi_2$.
The other contributions are given by simply 
replacing $X_k$, $Y_k$ with $X_k$, $Z_k$ for $\xi_1$, $\xi_3$
and with $Y_k$, $Z_k$ for $\xi_2$, $\xi_3$.

Denote
$D_k \equiv X_k-Y_k$ and $R \equiv e^{i D_1}$.
Again, free correlators are omitted.

\subsubsection*{$\alpha_0 \beta_0$ regularization}
\be
-\log \left(e^{-i X_1} -e^{-i Y_1} \right)
\ee

\subsubsection*{$\alpha_0 \beta_1$ regularization}
\be
\frac{R}{(1-R)^3} \Bigg[
i (R-1)^2 Y_2-\frac12 i D_2^2 (R+1)+D_3 (R-1)
 \Bigg]
\ee
For $\alpha_1 \beta_0$ regularization,
interchange $X_k \leftrightarrow Y_k$.

\subsubsection*{$\alpha_1 \beta_1$ regularization}
\begin{align}
&\frac{R^3}{(1-R)^6} \Bigg[\frac{11 D_2^4}{4}-3 D_2^3+6 D_2^2 X_2+\frac{1}{R^2}\bigg(\frac{D_2^4}{24}+\frac{D_2^3}{2}+D_2^2 \left(-X_2-\frac{1}{2} (i D_3)\right)
\nn \\
&+D_2 (-3 i D_3-D_4-X_2+4 i X_3)-\frac{D_3^2}{2}+2 i D_3 X_2-3 D_4+i D_5+X_2^2+6 X_4\bigg)
\nn \\
&+R^2 \bigg(\frac{D_2^4}{24}+\frac{D_2^3}{2}+\frac{1}{2} i D_2^2 (D_3+2 i X_2)+D_2 (3 i D_3-D_4-X_2-4 i X_3)-\frac{D_3^2}{2}
\nn \\
&-2 i D_3 X_2-3 D_4-i D_5+X_2^2+6 X_4\bigg)
+R \bigg(\frac{13 D_2^4}{12}+D_2^3+D_2^2 (-2 X_2+5 i D_3)
\nn \\
&+D_2 (-6 i D_3-2 D_4+4 X_2+8 i X_3)-D_3^2+4 i D_3 X_2+12 D_4+2 i D_5-4 X_2^2-24 X_4\bigg)
\nn \\
&+\frac1{R}\bigg(\frac{13 D_2^4}{12}+D_2^3+D_2^2 (-2 X_2-5 i D_3)+D_2 (6 i D_3-2 D_4+4 X_2-8 i X_3)-D_3^2
\nn \\
&-4 i D_3 X_2+12 D_4-2 i D_5-4 X_2^2-24 X_4\bigg)+6 D_2 (D_4-X_2)
\nn \\
&+3 \left(D_3^2+2 \left(-3 D_4+X_2^2+6 X_4\right)\right) \Bigg]
\end{align}

\subsubsection*{$\alpha_2 \beta_1$ regularization}
\begin{align}
&-\frac{R}{(1-R)^9}
\Bigg[i (R+1) X_2^3 (R-1)^6+12 i (R+1) X_3^2 (R-1)^6+15 i (R+1) X_6 (R-1)^6 
\nn \\
&+4 \left(3 D_2 \left(R^2+4 R+1\right)-(R-1)^2\right) X_5 (R-1)^5-\frac{1}{2} i \Big(3 \left(R^3+11 R^2+11 R+1\right) D_2^2
\nn \\
&+2 (R-1)^2 (R+1) D_2+6 D_3 i \left(R^3+3 R^2-3 R-1\right)\Big) X_2^2 (R-1)^4
\nn \\
&-\frac{3}{2} i \Big(3 \left(R^3+11 R^2+11 R+1\right) D_2^2+4 (R-1)^2 (R+1) D_2
\nn \\
&+6 D_3 i \left(R^3+3 R^2-3 R-1\right)\Big) X_4 (R-1)^4-\Big(\left(R^4+26 R^3+66 R^2+26 R+1\right) D_2^3
\nn \\
&+5 (R-1)^2 \left(R^2+4 R+1\right) D_2^2+2 (R-1) \left(3 i D_3 \left(R^3+11 R^2+11 R+1\right)-(R-1)^3\right) D_2
\nn \\
&+2 i (R-1)^2 \left(5 D_3 \left(R^2-1\right)+3 D_4 i \left(R^2+4 R+1\right)\right)\Big) X_3 (R-1)^3
\nn \\
&+\frac{1}{720} \bigg[-i \left(R^7+247 R^6+4293 R^5+15619 R^4+15619 R^3+4293 R^2+247 R+1\right) D_2^6
\nn \\
&-42 i (R-1)^2 \left(R^5+57 R^4+302 R^3+302 R^2+57 R+1\right) D_2^5
\nn \\
&+30 (R-1) \big(D_3 \left(R^6+120 R^5+1191 R^4+2416 R^3+1191 R^2+120 R+1\right)
\nn \\
&-8 i (R-1)^3 \left(R^3+11 R^2+11 R+1\right)\big) D_2^4+120 (R-1)^2 \Big(7 D_3 \big(R^5+25 R^4+40 R^3
\nn \\
&-40 R^2-25 R-1\big)+D_4 i \left(R^5+57 R^4+302 R^3+302 R^2+57 R+1\right)\Big) D_2^3
\nn \\
&+180 (R-1)^2 \Big(16 D_3 \left(R^2+4 R+1\right) (R-1)^3+2 i 
\big(7 D_4 \left(R^4+10 R^3-10 R-1\right)
\nn \\
&+D_5 i \left(R^4+26 R^3+66 R^2+26 R+1\right)\big) (R-1)+D_3^2 i \big(R^5+57 R^4+302 R^3+302 R^2
\nn \\
&+57 R+1\big)\Big) D_2^2+360 i (R-1)^3 \Big(7 \left(R^4+10 R^3-10 R-1\right) D_3^2 
\nn \\
&+2 D_4 i \left(R^4+26 R^3+66 R^2+26 R+1\right) D_3+2 (R-1) \big(8 D_4 (R+1) (R-1)^2
\nn \\
&+i \left(7 D_5 \left(R^3+3 R^2-3 R-1\right)+D_6 i \left(R^3+11 R^2+11 R+1\right)\right)\big)\Big) D_2
\nn \\
&-120 (R-1)^3 \bigg(\left(R^4+26 R^3+66 R^2+26 R+1\right) D_3^3-24 i (R-1)^3 (R+1) D_3^2
\nn \\
&+6 (R-1) \left(7 D_4 \left(R^3+3 R^2-3 R-1\right)+D_5 i \left(R^3+11 R^2+11 R+1\right)\right) D_3
\nn \\
&+3 (R-1) \Big(i \left(R^3+11 R^2+11 R+1\right) D_4^2+2 (R-1) \big(8 D_5 (R-1)^2+7 D_6 i \left(R^2-1\right)
\nn \\
&-D_7 \left(R^2+4 R+1\right)\big)\Big)\bigg)\bigg]
+X_2 \bigg[18 i (R+1) X_4 (R-1)^6+2 \Big(6 D_2 \left(R^2+4 R+1\right)
\nn \\
&-(R-1)^2\Big) X_3 (R-1)^5+\frac{1}{24} i \bigg(3 \left(R^5+57 R^4+302 R^3+302 R^2+57 R+1\right) D_2^4
\nn \\
&+32 (R-1)^2 \left(R^3+11 R^2+11 R+1\right) D_2^3+36 D_3 i \big(R^5+25 R^4+40 R^3-40 R^2
\nn \\
&-25 R-1\big) D_2^2+24 i (R-1)^2 \Big(8 D_3 \left(R^3+3 R^2-3 R-1\right)
\nn \\
&+3 D_4 i \left(R^3+11 R^2+11 R+1\right)\Big) D_2-12 (R-1)^2 \Big(3 \left(R^3+11 R^2+11 R+1\right) D_3^2
\nn \\
&+2 (R-1) \left(8 D_4 \left(R^2-1\right)+3 D_5 i \left(R^2+4 R+1\right)\right)\Big)\bigg) (R-1)^2\bigg] \Bigg]
\end{align}
For $\alpha_1 \beta_2$ regularization,
interchange $X_k \leftrightarrow Y_k$.

\subsubsection*{$\alpha_2 \beta_2$ regularization}
\begin{align}
-&\frac{R}{40320 (1-R)^{12}}
\Bigg[-\big(R^{10}+2036 R^9+152637 R^8+2203488 R^7+9738114 R^6+15724248 R^5
\nn \\
&+9738114 R^4+2203488 R^3+152637 R^2+2036 R+1\big) D_2^8
-112 (R-1)^2 \big(R^8+502 R^7
\nn\\
&+14608 R^6+88234 R^5+156190 R^4+88234 R^3+14608 R^2+502 R+1\big) D_2^7
\nn\\
&+56 (R-1) \Big(2 (R-1) \big((2 X_2-25) R^8+2 (502 X_2-1475) R^7+8 (3652 X_2-2975) R^6
\nn\\
&+2 (88234 X_2-1925) R^5+10 (31238 X_2+6125) R^4+2 (88234 X_2-1925) R^3
\nn\\
&+8 (3652 X_2-2975) R^2+2 (502 X_2-1475) R+2 X_2-25\big)-i D_3 \big(R^9+1013 R^8+47840 R^7
\nn\\
&+455192 R^6+1310354 R^5+1310354 R^4+455192 R^3+47840 R^2+1013 R+1\big)\Big) D_2^6
\nn\\
&+336 (R-1)^2 \Big(-14 D_3 i \left(R^8+246 R^7+4046 R^6+11326 R^5-11326 R^3-4046 R^2-246 R-1\right)
\nn\\
&+D_4 \left(R^8+502 R^7+14608 R^6+88234 R^5+156190 R^4+88234 R^3+14608 R^2+502 R+1\right)
\nn\\
&+2 (R-1) \big((15 X_2+4 i X_3-20) R^7+(1785 X_2+988 i X_3-460) R^6
\nn\\
&+9 (1785 X_2+1908 i X_3+20) R^5+(18375 X_2+62476 i X_3+1900) R^4
\nn\\
&+(-18375 X_2+62476 i X_3-1900) R^3-9 (1785 X_2-1908 i X_3+20) R^2
\nn\\
&+(-1785 X_2+988 i X_3+460) R-15 X_2+4 i X_3+20\big)\Big) D_2^5
\nn \\
&+840 (R-1)^2 \bigg(\big(R^8+502 R^7
+14608 R^6+88234 R^5+156190 R^4+88234 R^3+14608 R^2
\nn \\
&+502 R+1\big) D_3^2+4 i \left(R^2-1\right) \big((2 X_2-25) R^6+6 (82 X_2-225) R^5+(8094 X_2-3375) R^4
\nn \\
&+4 (5786 X_2+2375) R^3+(8094 X_2-3375) R^2+6 (82 X_2-225) R+2 X_2-25\big) D_3
\nn \\
&+2 (R-1) \Big(14 D_4 \left(R^7+119 R^6+1071 R^5+1225 R^4-1225 R^3-1071 R^2-119 R-1\right)
\nn \\
&+D_5 i \left(R^7+247 R^6+4293 R^5+15619 R^4+15619 R^3+4293 R^2+247 R+1\right)
\nn \\
&-2 (R-1) \big(-20 \left(R^4+26 R^3+66 R^2+26 R+1\right) X_2 (R-1)^2+3 \big(R^6+120 R^5+1191 R^4
\nn \\
&+2416 R^3+1191 R^2+120 R+1\big) X_2^2+6 \big((R^6+120 R^5+1191 R^4+2416 R^3+1191 R^2
\nn \\
&+120 R+1) X_4-4 i \left(R^6+56 R^5+245 R^4-245 R^2-56 R-1\right) X_3\big)\big)\Big)\bigg) D_2^4
\nn \\
&+6720 (R-1)^3 \bigg(7 \left(R^7+119 R^6+1071 R^5+1225 R^4-1225 R^3-1071 R^2-119 R-1\right) D_3^2
\nn \\
&+i \Big(D_4 \left(R^7+247 R^6+4293 R^5+15619 R^4+15619 R^3+4293 R^2+247 R+1\right)
\nn \\
&+2 (R-1) \big((15 X_2+4 i X_3-20) R^6+40 (21 X_2+12 i X_3-4) R^5
\nn \\
&+(3675 X_2+4764 i X_3+380) R^4+9664 i X_3 R^3+(-3675 X_2+4764 i X_3-380) R^2
\nn \\
&-40 (21 X_2-12 i X_3-4) R-15 X_2+4 i X_3+20\big)\Big) D_3-(R-1) \Big(18 X_2^2 R^6+D_6 R^6
\nn \\
&-44 i X_3 R^6+24 i X_2 X_3 R^6+54 X_4 R^6+16 i X_5 R^6+432 X_2^2 R^5+120 D_6 R^5-352 i X_3 R^5
\nn \\
&+1344 i X_2 X_3 R^5+1296 X_4 R^5+896 i X_5 R^5+270 X_2^2 R^4+1191 D_6 R^4+836 i X_3 R^4
\nn \\
&+5880 i X_2 X_3 R^4+810 X_4 R^4+3920 i X_5 R^4-1440 X_2^2 R^3+2416 D_6 R^3-4320 X_4 R^3
\nn 
\end{align}

\begin{align}
&+270 X_2^2 R^2+1191 D_6 R^2-836 i X_3 R^2-5880 i X_2 X_3 R^2+810 X_4 R^2-3920 i X_5 R^2
\nn \\
&+432 X_2^2 R+120 D_6 R+352 i X_3 R-1344 i X_2 X_3 R+1296 X_4 R-896 i X_5 R+18 X_2^2+D_6
\nn \\
&-14 i D_5 \left(R^6+56 R^5+245 R^4-245 R^2-56 R-1\right)+2 D_4 \big((2 X_2-25) R^6+120 (2 X_2-5) R^5
\nn \\
&+3 (794 X_2-125) R^4+16 (302 X_2+125) R^3+3 (794 X_2-125) R^2+120 (2 X_2-5) R
\nn \\
&+2 X_2-25\big)+44 i X_3-24 i X_2 X_3+54 X_4-16 i X_5\Big)\bigg) D_2^3+3360 (R-1)^3 \bigg(i (R^7+247 R^6
\nn \\
&+4293 R^5+15619 R^4+15619 R^3+4293 R^2+247 R+1) D_3^3-6 (R-1) \big((2 X_2-25) R^6
\nn \\
&+120 (2 X_2-5) R^5+3 (794 X_2-125) R^4+16 (302 X_2+125) R^3+3 (794 X_2-125) R^2
\nn \\
&+120 (2 X_2-5) R+2 X_2-25\big) D_3^2+6 i (R-1) \Big(14 D_4 (R^6+56 R^5+245 R^4-245 R^2
\nn \\
&-56 R-1)+D_5 i \left(R^6+120 R^5+1191 R^4+2416 R^3+1191 R^2+120 R+1\right) 
\nn \\
&-2 (R-1) \big(-20 \left(R^3+11 R^2+11 R+1\right) X_2 (R-1)^2+3 (R^5+57 R^4+302 R^3+302 R^2
\nn \\
&+57 R+1) X_2^2-6 i \big(4 \left(R^5+25 R^4+40 R^3-40 R^2-25 R-1\right) X_3+i (R^5+57 R^4+302 R^3
\nn \\
&+302 R^2+57 R+1) X_4\big)\big)\Big) D_3-3 (R-1) \Big((R^6+120 R^5+1191 R^4+2416 R^3+1191 R^2
\nn \\
&+120 R+1) D_4^2+4 (R-1) \big((15 X_2+4 i X_3-20) R^5+(375 X_2+228 i X_3-20) R^4
\nn \\
&+8 (75 X_2+151 i X_3+20) R^3-8 (75 X_2-151 i X_3+20) R^2+(-375 X_2+228 i X_3+20) R
\nn \\
&-15 X_2+4 i X_3+20\big) D_4+2 (R-1) \big(-4 X_2^3 R^5+2 X_2^2 R^5-24 X_3^2 R^5+D_7 i R^5+4 i X_3 R^5
\nn \\
&+48 i X_2 X_3 R^5-36 X_2 X_4 R^5+36 X_4 R^5+48 i X_5 R^5-20 X_6 R^5-100 X_2^3 R^4+2 X_2^2 R^4
\nn \\
&-600 X_3^2 R^4+57 D_7 i R^4-12 i X_3 R^4+432 i X_2 X_3 R^4-900 X_2 X_4 R^4+36 X_4 R^4+432 i X_5 R^4
\nn \\
&-500 X_6 R^4-160 X_2^3 R^3-16 X_2^2 R^3-960 X_3^2 R^3+302 D_7 i R^3+8 i X_3 R^3-480 i X_2 X_3 R^3
\nn \\
&-1440 X_2 X_4 R^3-288 X_4 R^3-480 i X_5 R^3-800 X_6 R^3+160 X_2^3 R^2+16 X_2^2 R^2+960 X_3^2 R^2
\nn \\
&+302 D_7 i R^2+8 i X_3 R^2-480 i X_2 X_3 R^2+1440 X_2 X_4 R^2+288 X_4 R^2-480 i X_5 R^2+800 X_6 R^2
\nn \\
&+100 X_2^3 R-2 X_2^2 R+600 X_3^2 R+57 D_7 i R-12 i X_3 R+432 i X_2 X_3 R+900 X_2 X_4 R-36 X_4 R
\nn \\
&+432 i X_5 R+500 X_6 R+4 X_2^3-2 X_2^2+24 X_3^2+D_7 i+14 D_6 (R^5+25 R^4+40 R^3-40 R^2
\nn \\
&-25 R-1)+2 D_5 i (R+1) \big((2 X_2-25) R^4+8 (14 X_2-25) R^3+6 (82 X_2+75) R^2
\nn\\
&+8 (14 X_2-25) R+2 X_2-25\big)+4 i X_3+48 i X_2 X_3+36 X_2 X_4-36 X_4+48 i X_5+20 X_6\big)\Big)\bigg) D_2^2
\nn \\
&+6720 (R-1)^4 \bigg(14 i \left(R^6+56 R^5+245 R^4-245 R^2-56 R-1\right) D_3^3-3 \Big(D_4 (R^6+120 R^5
\nn \\
&+1191 R^4+2416 R^3+1191 R^2+120 R+1)+2 (R-1) \big((15 X_2+4 i X_3-20) R^5
\nn \\
&+(375 X_2+228 i X_3-20) R^4+8 (75 X_2+151 i X_3+20) R^3-8 (75 X_2-151 i X_3+20) R^2
\nn \\
&+(-375 X_2+228 i X_3+20) R-15 X_2+4 i X_3+20\big)\Big) D_3^2-6 i (R-1) \Big(2 D_4 (R+1) \big((2 X_2
\nn \\
&-25) R^4+8 (14 X_2-25) R^3+6 (82 X_2+75) R^2+8 (14 X_2-25) R+2 X_2-25\big)
\nn \\
&-i \big(14 D_5 \left(R^5+25 R^4+40 R^3-40 R^2-25 R-1\right)+D_6 i (R^5+57 R^4+302 R^3+302 R^2
\nn \\
&+57 R+1)+2 (R-1) (27 i X_4 R^4-8 X_5 R^4+270 i X_4 R^3-208 X_5 R^3-528 X_5 R^2
\nn \\
&-270 i X_4 R-208 X_5 R+9 i \left(R^4+10 R^3-10 R-1\right) X_2^2+22 (R-1)^2 \left(R^2+4 R+1\right) X_3
\nn 
\end{align}

\begin{align}
&-12 \left(R^4+26 R^3+66 R^2+26 R+1\right) X_2 X_3-27 i X_4-8 X_5)\big)\Big) D_3
\nn \\
&-6 (R-1) \Big(7 \left(R^5+25 R^4+40 R^3-40 R^2-25 R-1\right) D_4^2+\big(i D_5 (R^5+57 R^4+302 R^3
\nn \\
&+302 R^2+57 R+1)-2 (R-1) (-20 \left(R^2+4 R+1\right) X_2 (R-1)^2
+3 (R^4+26 R^3
\nn \\
&+66 R^2+26 R+1) X_2^2+6 ((R^4+26 R^3+66 R^2+26 R+1) X_4
\nn \\
&-4 i (R^4+10 R^3-10 R-1) X_3))\big) D_4+i (R-1) \big(D_5 ((30 X_2+8 i X_3-40) R^4
\nn \\
&+4 (75 X_2+52 i X_3+20) R^3+528 i X_3 R^2-4 (75 X_2-52 i X_3+20) R-30 X_2+8 i X_3+40)
\nn \\
&+i \big(2 X_2^3 R^4+24 X_3^2 R^4+D_8 R^4+24 i X_2^2 X_3 R^4+4 i X_2 X_3 R^4+36 X_2 X_4 R^4+72 i X_3 X_4 R^4
\nn \\
&-40 i X_5 R^4+48 i X_2 X_5 R^4+30 X_6 R^4+24 i X_7 R^4+4 X_2^3 R^3+48 X_3^2 R^3+26 D_8 R^3
\nn \\
&+240 i X_2^2 X_3 R^3-8 i X_2 X_3 R^3+72 X_2 X_4 R^3+720 i X_3 X_4 R^3+80 i X_5 R^3+480 i X_2 X_5 R^3
\nn \\
&+60 X_6 R^3+240 i X_7 R^3-12 X_2^3 R^2-144 X_3^2 R^2+66 D_8 R^2-216 X_2 X_4 R^2-180 X_6 R^2
\nn \\
&+4 X_2^3 R+48 X_3^2 R+26 D_8 R-240 i X_2^2 X_3 R+8 i X_2 X_3 R+72 X_2 X_4 R-720 i X_3 X_4 R
\nn \\
&-80 i X_5 R-480 i X_2 X_5 R+60 X_6 R-240 i X_7 R+2 X_2^3+24 X_3^2+D_8
\nn \\
&-14 i D_7 \left(R^4+10 R^3-10 R-1\right)+2 D_6 ((2 X_2-25) R^4+(52 X_2-50) R^3+6 (22 X_2+25) R^2
\nn \\
&+(52 X_2-50) R+2 X_2-25)-24 i X_2^2 X_3-4 i X_2 X_3+36 X_2 X_4-72 i X_3 X_4+40 i X_5
\nn \\
&-48 i X_2 X_5+30 X_6-24 i X_7\big)\big)\Big)\bigg) D_2
-1680 (R-1)^4 \bigg((R^6+120 R^5+1191 R^4+2416 R^3
\nn \\
&+1191 R^2+120 R+1) D_3^4+8 i \left(R^2-1\right) \big((2 X_2-25) R^4+8 (14 X_2-25) R^3
\nn \\
&+6 (82 X_2+75) R^2+8 (14 X_2-25) R+2 X_2-25\big) D_3^3+12 (R-1) \Big(14 D_4 (R^5+25 R^4
\nn \\
&+40 R^3-40 R^2-25 R-1)+D_5 i \left(R^5+57 R^4+302 R^3+302 R^2+57 R+1\right)
\nn \\
&-2 (R-1) \big(-20 \left(R^2+4 R+1\right) X_2 (R-1)^2+3 \left(R^4+26 R^3+66 R^2+26 R+1\right) X_2^2
\nn \\
&+6 \left(\left(R^4+26 R^3+66 R^2+26 R+1\right) X_4-4 i \left(R^4+10 R^3-10 R-1\right) X_3\right)\big)\Big) D_3^2
\nn \\
&+12 i (R-1) \Big(\left(R^5+57 R^4+302 R^3+302 R^2+57 R+1\right) D_4^2
+4 (R-1) ((15 X_2+4 i X_3
\nn \\
&-20) R^4+2 (75 X_2+52 i X_3+20) R^3+264 i X_3 R^2-2 (75 X_2-52 i X_3+20) R-15 X_2
\nn \\
&+4 i X_3+20) D_4+2 i (R-1) \big(2 D_5 ((2 X_2-25) R^4+(52 X_2-50) R^3+6 (22 X_2+25) R^2
\nn \\
&+(52 X_2-50) R+2 X_2-25)-i \big(14 D_6 \left(R^4+10 R^3-10 R-1\right)+D_7 i (R^4+26 R^3+66 R^2
\nn \\
&+26 R+1)-2 (R-1) (2 \left(R^3+11 R^2+11 R+1\right) X_2^3-(R-1)^2 (R+1) X_2^2
\nn \\
&+6 \left(3 \left(R^3+11 R^2+11 R+1\right) X_4-4 i \left(R^3+3 R^2-3 R-1\right) X_3\right) X_2+2 (-i X_3 (R-1)^3
\nn \\
&-9 (R+1) X_4 (R-1)^2+6 \left(R^3+11 R^2+11 R+1\right) X_3^2-12 i R^3 X_5-36 i R^2 X_5+12 i X_5
\nn \\
&+36 i R X_5+5 R^3 X_6+55 R^2 X_6+55 R X_6+5 X_6))\big)\big)\Big) D_3-12 (R-1)^2 \Big(2 ((2 X_2-25) R^4
\nn \\
&+(52 X_2-50) R^3+6 (22 X_2+25) R^2+(52 X_2-50) R+2 X_2-25) D_4^2
\nn \\
&+2 \big(-14 D_5 i \left(R^4+10 R^3-10 R-1\right)+D_6 \left(R^4+26 R^3+66 R^2+26 R+1\right)
\nn \\
&+2 (R-1) (27 X_4 R^3+8 i X_5 R^3+81 X_4 R^2+88 i X_5 R^2-81 X_4 R+88 i X_5 R
\nn \\
&+9 \left(R^3+3 R^2-3 R-1\right) X_2^2-22 i (R-1)^2 (R+1) X_3+12 i \left(R^3+11 R^2+11 R+1\right) X_2 X_3
\nn 
\end{align}
\begin{align}
&-27 X_4+8 i X_5)\big) D_4+D_5^2 \left(R^4+26 R^3+66 R^2+26 R+1\right)
\nn \\
&+4 D_5 i (R-1) \left(-20 (R+1) X_2 (R-1)^2+3 \left(R^3+11 R^2+11 R+1\right) X_2^2 \right.
\nn \\
&\left. +6 \left(\left(R^3+11 R^2+11 R+1\right) X_4-4 i \left(R^3+3 R^2-3 R-1\right) X_3\right)\right)
\nn \\
&+2 (R-1) \big(-R^3 X_2^4-3 R^2 X_2^4+3 R X_2^4+X_2^4-36 R^3 X_4 X_2^2-108 R^2 X_4 X_2^2+108 R X_4 X_2^2
\nn \\
&+36 X_4 X_2^2-48 R^3 X_3^2 X_2-144 R^2 X_3^2 X_2+144 R X_3^2 X_2+48 X_3^2 X_2-60 R^3 X_6 X_2
\nn \\
&-180 R^2 X_6 X_2+180 R X_6 X_2+60 X_6 X_2+14 D_8 R^3+D_9 i R^3+42 D_8 R^2+11 D_9 i R^2
\nn \\
&-4 R^3 X_3^2+12 R^2 X_3^2-12 R X_3^2+4 X_3^2-54 R^3 X_4^2-162 R^2 X_4^2+162 R X_4^2+54 X_4^2-14 D_8
\nn \\
&+D_9 i-42 D_8 R+11 D_9 i R+2 D_7 i (R+1) \left((2 X_2-25) R^2+10 (2 X_2+5) R+2 X_2-25\right)
\nn \\
&+2 D_6 ((15 X_2+4 i X_3-20) R^3+(45 X_2+44 i X_3+60) R^2+(-45 X_2+44 i X_3-60) R
\nn \\
&-15 X_2+4 i X_3+20)-96 R^3 X_3 X_5-288 R^2 X_3 X_5+288 R X_3 X_5+96 X_3 X_5+80 R^3 X_6
\nn \\
&-240 R^2 X_6+240 R X_6-80 X_6-28 R^3 X_8-84 R^2 X_8+84 R X_8+28 X_8\big)\Big)\bigg) \Bigg]
\end{align}

\begin{center}
{\bf Acknowledgements}
\end{center}

The authors acknowledge the support of this work by  the National Natural 
Science Foundation of China under grant 11575119.


\end{document}